\documentclass[apj]{emulateapj}
\usepackage{apjfonts}

\def\lax {\ifmmode{_<\atop^{\sim}}\else{${_<\atop^{\sim}}$}\fi}  
\def\gax {\ifmmode{_>\atop^{\sim}}\else{${_>\atop^{\sim}}$}\fi}  
\def\gtorder{\mathrel{\raise.3ex\hbox{$>$}\mkern-14mu
             \lower0.6ex\hbox{$\sim$}}}

\def\etal{{\it et al.\/} } 

\begin{document}

\title{Correlations between X-ray Spectral Characteristics and\\
Quasi-Periodic Oscillations in Sco X-1}

\author{Charles F. Bradshaw\altaffilmark{1}, Lev Titarchuk\altaffilmark{2} and 
Sergey Kuznetsov\altaffilmark{3}}

\shorttitle{X-RAY SPECTRUM AND QPO IN SCO~X-1}

\shortauthors{BRADSHAW, TITARCHUK \& KUZNETSOV}

\journalinfo{The Astrophysical Journal, 663: 000-000, 2007, July 1}

\slugcomment{accepted 2007 March 14}

\altaffiltext{1}{The MITRE Corporation, McLean, VA 22102 and George Mason University Center for Earth Observing and Space Research, Fairfax, VA 22030; email:cbradshaw@mitre.org}
\altaffiltext{2}{George Mason University/Center for Earth Observing and Space Research, Fairfax, VA 22030; US Naval 
Research Laboratory, Code 7655, Washington, DC 20375-5352; and Goddard Space Flight Center, NASA, code 661, Greenbelt MD 20771; email:ltitarchuk@ssd5.nrl.navy.mil \& lev@milkyway.gsfc.nasa.gov}
\altaffiltext{3}{IGPP, University of California, Riverside, CA 92521;
Space Research Institute of Russian Academy of Sciences, Profsoyuznaya 
84/32, Moscow 117997, Russia; email:sergeyk@ucr.edu}

\begin{abstract}
Correlations between 1-10 Hz quasi-periodic oscillations (QPOs) and spectral power law index have been reported for black hole (BH) candidate sources and one neutron star source, 4U 1728-34.  An examination of QPO frequency and index relationships in Sco X-1  is reported herein.  We discovered that Sco X-1, representing Z-source groups, can be adequately modeled by a simple two-component model of Compton up-scattering  with a soft photon electron temperature of about 0.4 keV, plus an Iron K-line. The results show a strong correlation between spectral power law index and kHz QPOs.  Because Sco X-1 radiates  near the Eddington limit,  one can infer that the geometrical configuration of the Compton cloud (CC)  is quasi-spherical because of high radiation pressure in the CC.  Thus, we conclude  that the high Thomson optical depth of the Compton cloud, in the range of $\sim 5-6$ from the best-fit model parameters, is consistent with the neutron star's surface being obscured by material.  Moreover, a spin frequency of Sco X-1 is likely suppressed due to photon scattering off CC electrons.   Additionally,  we demonstrate how the power spectrum evolves when Sco X-1  transitions from the horizontal branch to the normal branch.

\end{abstract}


\section{Introduction}
Sco X-1 is a bright low-mass X-ray binary (LMXB) system and is considered a prototype for these systems.  Sco X-1 is classified as a Z source because it forms a three-branch Z-shape in its color diagrams. The three branches are designated as the horizontal branch, normal branch and flaring branch (HB, NB and FB), for the top, middle, and bottom of the Z, respectively. 
 
Z-sources can be divided into two groups: (1) Cygnus-like sources, e.g.,  Cyg X-2, GX5-1 and GX 340+0 and (2) Sco-like sources, e.g.,  Sco X-1, GX 17+2 and GX 349+2  \citep{kuu97}. The two groups differ in that the hardness-intensity and color diagrams in Sco-like sources show an almost vertical ``horizontal'' branch.  The differences between the two group's morphology in the color diagrams is not completely understood.

The standard Z-source model is a neutron star (NS) accreting from a secondary normal star such that the accretion rate increases monotonically from the horizontal branch to the flaring branch by about a factor of two \citep{has90b}. The standard model accretion disk for these sources is optically thick but geometrically thin.  LMXBs manifest X-ray quasi-periodic oscillations (QPOs) from a few Hz to kHz and both the frequency and amplitude of the QPOs vary with color diagram position and, presumably, accretion rate.  

Observed and predicted LMXB oscillations have been fitted by a single model by Titarchuk et al. (2001a) in which a geometrically thin accretion disk in the horizontal branch changed to nearly spherical accretion in the normal branch.  The 6-20 Hz normal-branch oscillator (NBO) was identified with acoustic oscillations of a spherical shell around the neutron star on the order of one neutron star radius. Titarchuk \& Shaposhnikov (2005), hereafter TS05, suggested that black hole (BH) and NS binaries could be observationally distinguished by the correlation between the spectral indices and quasi-periodic oscillations in the binary X-ray spectra.  They reported that there was a correlation with the X-ray state and QPOs in the spectra of the neutron star binary 4U 1728-34, a low luminosity atoll source.  We postulate, that in high-luminosity LMXBs, like Sco X-1 and Cyg X-2, such correlations should occur if the physical processes producing the spectra are similar. 

\section{Observations}

Rossi X-ray Timing Explorer (RXTE) observations, during four epochs in 1997 and 1998 and nearly continuously for three days in 1999, were used herein to examine correlations between QPOs and photon spectral index.  The correlations were then compared with previously reported observations of Cyg X-1 \citep{sha06} and 4U 1728-34 (TS05).   A description of the observation data used in this paper was previously reported in Bradshaw \etal (2003).  The 1997 and 1998 data were observed in RXTE epoch 3.  The 1999 data were observed in RXTE epoch 4.  The different epochs and pointing parameters caused compression of the RXTE energy channels in 1999 when compared to the earlier epochs. The RXTE pointing offset was 24 arc minutes in the 1999 epoch and 12 arc minutes during the February 28, 1998 observation.  Due to different instrument channelization and pointing offsets, there was a shift in the hard color ratio (HCR) between the 1999 epochs and earlier epochs that required normalization before adequate comparisons between the epochs could be undertaken.  

To normalize all observations to a common HCR, we developed a normalized hardness color ratio (NHCR) for comparing HCR and QPO frequencies among all epochs based on the position of a ``${\rm HCR_{gap}}$'' seen in RXTE color-intensity diagrams \citep[Figure 1]{bra03}.  This gap HCR value was assumed to be a constant relative value for all observations at the bottom of the horizontal branch, providing a clear marker for a standard accretion rate.  Thus, the NHCR is produced by:  $1-(\rm HCR_{gap} - {\rm HCR_{observed}})$, where  ${\rm HCR_{gap}=0.90}$ for the 1997 and 1998 epochs and ${\rm HCR_{gap}}=0.87 $ for the 1999 epoch.  The NHCR yields a horizontal branch to normal branch transition equal to one and does not assume either a maximum or minimum for the observed HCR. The NHCR provides an accurate mechanism for determining the X-ray state across all Sco X-1 epochs.

\section{Power spectrum and QPO identification}

\begin{figure}[tb]
\epsscale{1.0}
\plotone{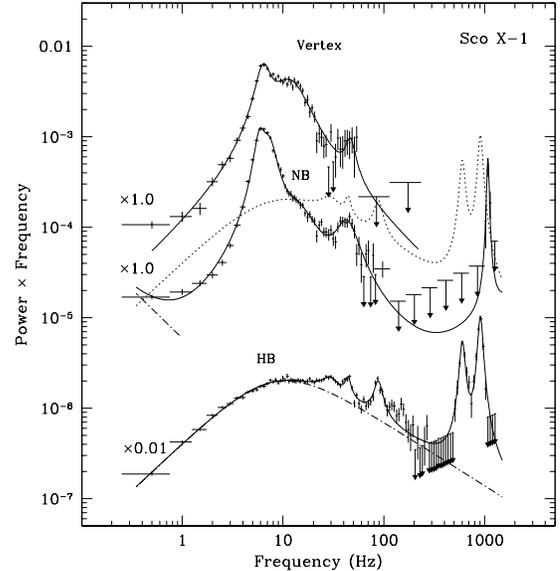}
\caption{$\nu \: \times$ power diagram of Sco X-1 while at the normal-flaring branch vertex, in the middle of the normal branch and in the horizontal branch. The dotted line shows the $\times$1.0 position of the horizontal branch $\nu \: \times$ power diagram.  The dash-dotted  lines depict a continuum fit to the HB and NB power spectra. 
}
\label{figure1}
\end{figure}

Figure \ref{figure1} illustrates the changes in the power spectral density (PSD) of the NB, HB and NB-HB vertex (hereafter Vertex), respectively.  Strong QPO features are clearly seen in each state's plotted $\nu \: \times$PSD. Also seen in Figure \ref{figure1}, the $\nu\:\times$PSD of both the NB and Vertex show a prominent $\sim 6$ Hz QPO. All $\nu\:\times$PSDs show a 40 Hz QPO but only the Vertex $\nu\:\times$PSD shows no kHz QPOs.

The transient nature  of kHz QPOs is presumably related to the  density profile of  the innermost part of the source. At very high mass accretion rates, there should be a density profile inversion that leads to (Raleigh-Taylor) instabilities of the QPOs  \citep{tit03}.  Another reason for the absence of kHz QPOs at the Vertex (softest state) is a smearing out of kHz QPOs in the wind \citep{tit02}. In \S \ref{wind}, we show that at NHCR$<0.9$ (Vertex)  a strong  K${_\alpha}-$ line emission is detected.  Also in \S \ref{wind} we demonstrate the wind is an unavoidable consequence of a high matter supply to the compact object (NS or BH) from the companion star (see more details in Titarchuk et al. 2007, hereafter TSA07). Laming \& Titarchuk (2004), hereafter LT04, present compelling arguments that a strong K${_\alpha -}$line can be formed in a powerful wind.  
    
QPO values reported herein were obtained by:  (1) Performing FFTs on the RXTE's 250 $\mu$s data, modeling the background FFT spectrum with either a constant or power law function, depending on the QPO's continuum background, (2)  modeling the QPO with a Lorentzian shape, and (3) calculating the root-mean-square (RMS) power percentage of the Lorentzian QPOs modeled.  The dash-dotted lines in Figure \ref{figure1} depict a continuum fit to the HB and NB power spectra. At the Vertex, the PSD continuum is very weak.

Each identified kHz QPO was modeled with a Lorentz profile plus a constant, where the constant was used to remove noise above $\sim$ 100 Hz.  The noise in this region of  the spectrum is flat and can be easily modeled with a constant.  Whenever an observed QPO could not be modeled with a physically reasonable error, the QPO was discarded.   
 
\section{Results}
\subsection{X-ray Spectra}

\begin{figure}[tb]
\epsscale{1.0}
\plotone{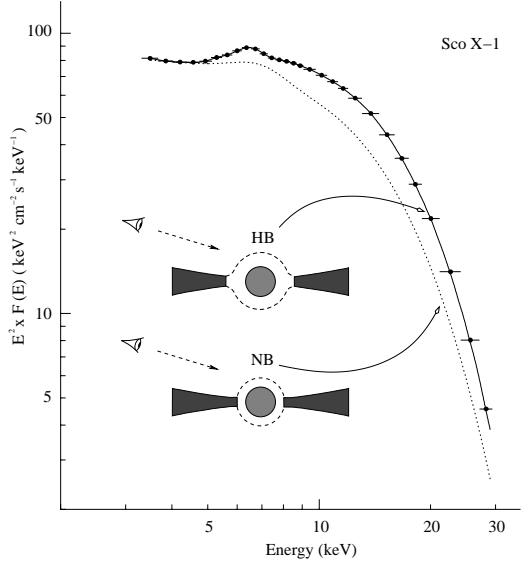}
\caption{The solid line is a CompTT + Fe K-line model fit of an observation when Sco X-1 was in the horizontal branch (HB). The dashed line is a CompTT + Fe K$_{\alpha}$-line model fit of an observation when Sco X-1 was in the normal branch (NB). Embedded diagrams illustrate the suggested positions  of Compton cloud  with respect to the accretion disk. The disk  obscures  the lower hemisphere of Compton cloud during the NB stage  (lower cartoon) and the lower hemisphere is exposed to the observer during the HB  stage  (upper cartoon). 
 }
\label{figure2}
\end{figure}

All spectra were modeled at the energy range of 3-30 keV with an added systematic error of 1.5\%.  The modeling results produced an average reduced $\chi^{2}$ ($\chi^2$ divided by the degrees of freedom) $ \chi^{2}_{red} = 0.74$ across all OBSIDs using the FTOOLS software package \citep{bla95}. This suggests that the errors were  slightly over estimated but very close to reality.  The spectra were modeled in XSPEC with a two-component additive model consisting of a Comptonization (CompTT) and an Iron K$_\alpha -$ line (Gaussian) model (see Figure \ref{figure2}). Photon index $\Gamma$ ($\Gamma$ equals $\alpha +1$) values were developed from the modeling results, where $\alpha$ is defined by:
\begin{equation}
\alpha = -3/2 + \sqrt{9/4 + \gamma}, ~~~{\rm and}
\label{alpha}
\end{equation}
\begin{equation}
\gamma = \frac{\pi^{2}m_ec^{2}}{3 \left(\tau + \frac{2}{3}\right)^{2} kT }.
\end{equation}

The optical depth $\tau$ and electron temperature $kT$ of  the Compton cloud (hereafter CC) values were determined from the CompTT model parameters and $m_ec^{2} = 511$ keV.  Error bars for $ \Gamma $ were determined from the errors of $ \tau$ and $ kT $.  The soft photon temperature (T0) parameter of CompTT was frozen to a value of 0.4 keV.  The standard error of T0 was determined by increasing its value in small steps until the $ \chi^{2}_{red} $ value increased by one, yielding $0.4 \pm 0.07$ keV.

\subsection{The inferred geometry of Sco X-1's X-ray emission region} 

In Figure \ref{figure2}, the apparent X-ray intensity in the HB is seen to be about  1.5 times higher than that of the NB. In the Sco X-1 color intensity diagram \citep{bra03},  the intensity remains nearly constant when the HCR varies within the horizontal branch (where NHCR$>1.0$). This lack of dependence of the intensity on changes in spectral hardness has also been observed in X-ray bursters during an expansion stage. In the X-ray burst contraction stages, the color temperature of the emergent blackbody-like spectrum changes dramatically,  from 0.5 keV to 3 keV, while  the intensity remains almost a constant at the Eddington limit  (see Shaposhnikov \& Titarchuk 2004).  The distance to Sco X-1, found by  Bradshaw \etal (1999), and the measured X-ray intensity implies that Sco X-1 emits very close to the Eddington limit in the HB or NB.  However, the effect of   spectral softening for a constant intensity in LMXBs can be different from that observed in X-ray bursters. In NS accretion, the total intensity is a sum  of the energy released in the accretion disk, Compton cloud, and NS surface and the hardness of the Comptonization spectrum is determined primarily by its plasma temperature, which is dictated by the X-ray flux originating at the NS surface. 

In observations of the Z-source NB, intensity decreases as the spectrum becomes softer (see Bradshaw \etal 2003, Fig. 1). To explain this dependency of the intensity vs color (HCR), we suggest (see Shaposhnikov \& Titarchuk 2004) that the photosphere (CC) geometry  changes during a transition from the HB to NB. In the HB, the CC is not fully extended to the disk, exposing the upper hemisphere of the photosphere and some portion of the lower hemisphere to the Earth observer.  Consequently, the apparent HB intensity is higher than the NB intensity when the CC is pushed by the disk (see the diagrams in Fig. \ref{figure2}). The disk surface shields the lower hemisphere, and therefore, the observer cannot detect the lower hemisphere's flux. These intensity variations allow us to estimate the disk inclination angle of the X-ray emission region $i$. Following the prescription from Shaposhnikov \& Titarchuk (2004),  we find that $i\sim 50^{o}-60^{o}$, which is  consistent with the results reported by Fomalont \etal (2001). Figure \ref{figure2} illustrates how the emergent spectrum of the emission region,  intensity, and geometry, evolve from the HB to the NB.

\subsection{The correlation of kHz QPO frequencies with photon index}

\begin{figure}[tb]
\epsscale{1.0}
\plotone{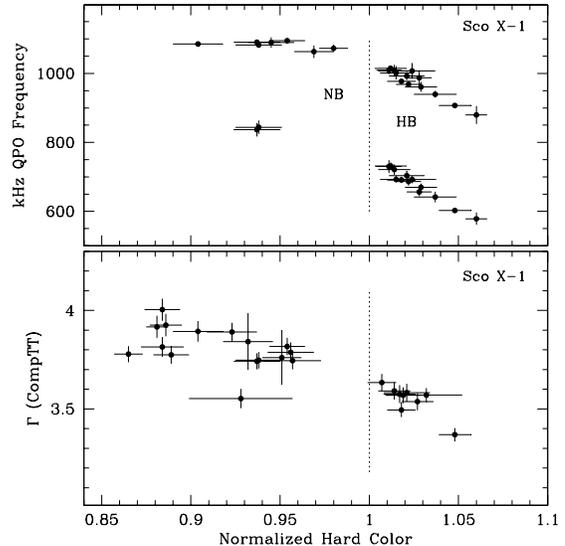}
\caption{High and Low kHz QPO frequencies and photon index $ \Gamma$ determined by the Comptonization model (CompTT') as a function of normalized hard color ratio.  The normalized hard color ratio of 1.0, shown by the dashed line, indicates the transition between the normal ($ < 1 $) and horizontal ($ > 1 $) branches. }
\label{figure3}
\end{figure}

Figure \ref{figure3} depicts the high-frequency (kHz) QPO and $ \Gamma $ variability  as a function of the NHCR.   The upper and lower kHz frequencies are anticorrelated with NHCR (see the upper and lower data points in the top panel, respectively) as is  $\Gamma$ (bottom panel). From this, we infer that there is a correlation of kHz QPO frequencies with $\Gamma$.  A similar correlation of QPOs with $\Gamma$ was found in 4U 1728-34 by TS05, where the QPO-index correlation was revealed when the source evolved from the low/hard state ($\Gamma\sim 2$) to the soft state ($\Gamma\gax 4$). In contrast, Sco X-1 always shows a soft spectrum and a QPO-index correlation as $\Gamma$ changes from $3.3$ to $4$.   The Sco X-1 photon indices  $\Gamma=3.3-4$ are very close to those in the soft state of 4U 1728-34. In both neutron stars, inferred $\Gamma$ of the Vertex, the softest state ($\sim 4$), is substantially higher than that of the soft state of black holes ($\sim 2.8$).

 \subsection{The correlation of the K$_{\alpha}-$ line EW and photon index $\Gamma$}
 \label{wind}

\begin{figure}[tb]
\epsscale{1.0}
\plotone{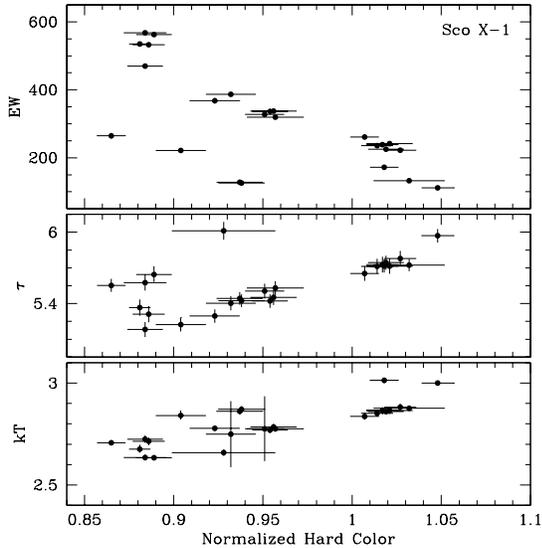}
\caption{CompTT parameters $kT$,  $\tau$ and the Gaussian component's equivalent width (EW) in eV   as a function of normalized hard color ratio.}
\label{figure4}
\end{figure}

Figure \ref{figure4} (upper panel) shows an anticorrelation of the equivalent Gaussian width (EW), a model of the K$_{\alpha}-$ line, with NHCR (photon index). The EW steadily increases as the state softens.  According to LT04,  the strength of the line (EW) should be related to the mass outflow rate  ($\dot M_{ofl}$). They show that, if the source emits at the Eddington limit, the optical depth of the wind 
$\tau_{W}$ is in the range of $0.5-1.0$ for  EW $\sim 0.5-1.0$ keV.  Titarchuk,  Shaposhnikov \& Arefiev (2007), (see also more details  in the next section and Appendix A)  show that the outflow mass rate  is directly related to the mass accretion rate. 

The optical depth of the wind $\tau_{W}$ is also related to the mass outflow rate (see  Eq. \ref{tau outflow2}) and is inversely proportional to the wind velocity and the effective plasma cross section $\sigma(R)$. In addition, $\tau_{W}$ is proportional to the relative width of the disk $H/R$.  When an oscillating X-ray emission passes through the intervening wind environment, the QPO frequencies are dampened, which contrasts with the increasing strength of the Iron K$_{\alpha}-$line as a function the optical depth of the wind $\tau_{W}$.  
  
\subsection{The inferred  optical depth of Sco X-1's CC  and its effect on detecting kHz QPOs}   
\label{optical_depth}

Our spectral model's high values for the CC's Thomson optical depth (5-6)  (see Figure \ref{figure4})  imply that high frequency QPOs originating at the NS are suppressed because of photon scattering off cloud electrons (Titarchuk 2002). This conclusion infers that the high (kHz) QPOs can be only detected by an Earth observer if they originate in the outer boundary of the CC (see the diagrams in Figure 1). Titarchuk et al.  (1998), Osherovich \& Titarchuk (1999) and Titarchuk (2003)  presented observational evidence that the lower kHz QPO is related to the Keplerian frequency and the upper kHz peak is related to the hybrid frequency at the outer boundary of the CC.  Moreover, the observed correlation of kHz frequencies with photon index (color or state) (Figure \ref{figure3}) indicates that the CC contracts when the source undergoes transitions to the soft state, producing the observed QPO frequency evolution.   

\subsection{The correlation (anticorrelation)  of the CC plasma temperature and the color ratio NHCR (photon index $\Gamma$)}

The relatively low $kT$  of the CC presented in Figure  \ref{figure4} indicates that the plasma is very close to equilibrium with the photon environment. As a result, the upscattering efficiency of the soft photons, presumably coming from  the NS surface and the disk,  in the cloud is low (high values of $\Gamma=3.3-4$ are a sign of inefficient upscattering) despite the high values of the optical depth.  
In Figure  \ref{figure4}, the CC plasma temperature $kT$ clearly decreases when the source evolves to softer states (from HB to NB), as indicated by the decreasing NHCR (increasing $\Gamma$ and decreasing Comptonization efficiency).

\section{Discussion}

In our X-ray data, we see correlations between spectral indices and kHz QPO frequencies. This is the first demonstration of such correlations in Z-source states. The spectral indices were inferred using a generic Comptonization  model  (CompTT), which is valid for any bounded configuration.  Because kHz QPOs tightly correlate with other timing features of the power spectra in Sco X-1 (break, HBO, and viscous frequencies,   see Titarchuk \etal 1999, 2001a), we can claim that we truly establish an index-frequency correlation in Sco X-1.
Because there is a large body of work on Z-source physics, it is natural to ask how {\it the results herein relate to  previous works. } 

In particular, Psaltis, Miller, \& Lamb (1995), hereafter PML95,   theoretically studied the emergent spectra from  NS sources in terms of a thermal Comptonization model.  Their main model parameters (excluding normalization) were the plasma temperature, the optical depth of the Compton cloud and the seed  photon temperature.  Our observed photon indices ($\Gamma$), shown in Figure 3, increase when the source evolves along the normal branch towards the flaring branch, which was also produced in PLM95's modeling, but our observed index values (3.5 to 4) are much greater than those in PLM95  (1.56 to 1.68).  Moreover, the observed values of  electron temperature ($kT_e=2.5-3$ keV) are much less than produced by PFM95, where $kT_e$ was higher than 6 keV. The observed values of $\tau$ also have a tendency to decrease along the NB (from 6.1 to 5.1) as in PLM05, but their theoretical value of $\tau$ (from 10 to 6.5) is  almost twice our observed values. The observed value of the soft temperature (0.4 keV) agrees with PLM05's theoretical values ($ <0.5$ keV).  

 The approach taken by Grebenev \& Sunyaev (2002), hereafter GS02, who computed the photon spectra formed in a boundary (transition layer)  near the surface of a neutron star  agrees perfectly with our observations of Sco X-1 (see Figures 2-4).  They included the frequency dependence of the free opacity and used a full description of the Comptonization process. This led to lower plasma temperatures and softer spectra than  was calculated in Popham \& Sunyaev (2001), hereafter PS01.   It is also worth noting that, PS01 recognize that their model predicts a decrease of kHz QPO frequency when the mass accretion increases,
which is the opposite of observed correlations between QPO frequencies with mass accretion rate (see van der Klis et al. 1996 for Sco X-1 and Homan et al. 2002).

Additionally, Revnivtsev \& Gilfanov (2006), hereafter RG06, studied the contribution of the boundary layer in the emergent X-spectrum of NS sources as a function of the Z-track position ($S_z$). They fit the spectra of various of NS sources  using two Comptonization (CompTT) components, one related to the NS emission and another to the boundary layer. TS05 were the first to publish this kind of analysis for the atoll source 4U 1728-34. They demonstrated that, in all the spectral states of 4U 1728-34, the emergent spectrum is a sum of two components related to two sources of soft seed photons; the NS and the disk. The neutron star and disk inject soft photons to the CC where soft photon upscattering in the CC hot plasma produce the resulting spectrum. We do not find a direct  contribution in  our data from upscattered NS photons, but our results find that the CC plasma temperature is very close to that found in RG06.  The CC temperature is likely dictated by strong emission from the NS, resulting from gravitational energy release at the its surface.

 The disk contribution, $L_d$,  in the total emission strongly depends on the position of the inner edge of the Keplerian disk ($R_{in}$), namely $L_d \propto R_{in}^{-1}$ (see Shakura \& Sunyaev 1973). 
  {\it Our observations provide insight into how the structure of 
the accretion disk and boundary (transition) layer vary together  with mass accretion  rate}.  We have found that, when the luminosity is very close to Eddington, the disk is no longer geometrically thin and can obscure the lower corona hemisphere from the observer by this puffed up structure (see detailed calculations  in PS01).  A fraction of the quasi-spherical corona is obscured from the observer by the geometrically thick disk.

In previous studies presented in the literature,  people have looked at the correlations between the position of a Z-source on its color-color diagram and either the spectra or the timing properties. Our paper is {\it the 
first to compare the kHz QPO frequency, the spectrum (using the power-law photon index $\Gamma$), and the iron flux} in a Z-source. For example, Homan et al. (2002), hereafter H02,  have undertaken an  extensive  frequency analysis where they analyzed the correlation of QPO frequencies along the Z-track. They used a ``rank number'' ($S_z$) that identified the position of selected observations along the Z-track. $S_z$ increased from 1 to 2 for HB/NB and NB/FB vertices, respectively. H02 found that when the source moves from the HB to the NB the color decreased, i.e., the spectrum becomes softer and this is when the HBO and kHz frequencies increase. Unfortunately, color value only has a phenomenological meaning and it is very difficult to interpret the QPO-color anticorrelation in physical terms.  

In contrast, our correlation of QPOs with the spectral index of the Comptonization has a direct physical interpretation. The anticorrelation of the index with the Comptonization parameter allows us to state that when the QPO frequency increases, the efficiency of Comptonization decreases. 
In Appendix  B, we show that  the Comptonization parameter $Y=\eta N_{sc}$, as a product of average fractional energy  change  per scattering  $\eta$ and the mean number of scattering $N_{sc}$ in the Compton cloud, is a reciprocal  of  spectral index $\alpha$ ($\alpha=\Gamma-1$), namely 
$Y\sim \alpha^{-1}$.   

 The physics of this  index-QPO correlation (phenomenon)  can be discussed in the framework of the transition layer around the NS and BH (Titarchuk, Lapidus \& Muslimov 1998, hereafter TLM98).
Our observations indicate that the origin of QPO frequencies detected from many NS and BH binaries (see the review by  McClintock, \&  Remillard 2006, hereafter MR06) is likely at the outer surface of the transition (boundary) layer between the Keplerian disk and neutron-star. The existence of the TL layer was derived, using first principles of fluid Physics  in the disk-like configuration, by TLM98.  If we look at the azimuthal momentum component of the equation of motion (see SS73, TLM98, \& PS01), we can  realize that this  angular moment exchange due to the disk viscosity  is a diffusion process. This implies that it is described by a differential equation of the second order where the only parameter is the Reynolds number of the flow, which is proportional to the mass accretion rate. Thus, with the three boundary conditions for the adjustment,  TLM98  found  that {\it the position (radius)  of the outer (adjustment) boundary $R_{adj}$ is determined by the Reynolds number only and moreover $R_{adj}$ decreases when  the mass accretion rate increases}. 

Titarchuk \& Osherovich  (1999) argue that  the QPO low and high frequencies $\nu_{low}$, $\nu_{h}$  are a frequency of volume (CC) oscillation and  an oscillation frequency  of the TL outer surface (shock), respectively.   Frequency $\nu_{low}$ as a volume oscillation frequency
$\nu_L\sim V_{MA}/R_{adj}$ is inversely proportional to the adjustment radius $R_{adj}$ 
(where $V_{MA}$ is a magneto-accoustic velocity) and the shock oscillation frequency 
$\nu_h\sim \nu_{\rm K}$  (where $\nu_{\rm K}$ is the Keplerian frequency at the adjustment radius) 
is inversely proportional to $R_{adj}^{3/2}$  [see Titarchuk et al. (2001b) for details].
Thus, we conclude that the CC contracts when the QPO frequency and index increases, moving the transition layer closer to the NS surface.  

The correlation between the spectral index $\Gamma$, QPOs and the mass accretion rate is  a well established observational fact (see TS05, MR06).  From a physical perspective, when mass accretion rate increases, the soft disk and NS photon flux increases.  This cools the plasma temperature $T$ of the transition layer (CC), decreasing the efficiency of Comptonization  $Y\sim (4kT/m_ec^2)\tau^2$ (see Appendix B).   Our analysis of the Sco X-1 observations shows that the CC temperature is very close to the color temperature of the NS photons, i.e. about 2.4-2.7 keV.  

Using these facts, and the index anticorrelation with the Comptonization (upscattering efficiency) parameter, $\alpha\sim Y^{-1}$,  {\it we conclude again that the index and QPO frequencies increase with the mass accretion rate, implying that the Compton cloud (TL),  which is probably the origin of the observed Comptonized X-ray emission,  contracts when the mass accretion rate increases.} 

In addition to the Compton spectra and QPOs, our observations also show a strong wind is launched from the disk  when the  mass supply to the disk is very high.  Recently   TSA07 demonstrate that a powerful wind may occur in X-ray  sources, like Sco X-1, Cyg X-2, and Cyg X-1,  which are characterized by a high matter supply from the companion. The wind is presumably a result of disk overflow when the mass accretion rate in the given annulus ($R, ~R+dR$) exceeds the local Eddington critical value $\dot M_d^{crit}(R)$ (see Appendix B and TSA07). 
From  the formula (\ref{tau outflow2}), given in the Appendix A  we conclude that the Thomson optical depth of the wind (outflow)  $\tau_W\gax 1$, for typical values of a  half-thickness of the disk $H\sim 0.1 R$, the effective plasma cross section $\sigma(R)\sim 5\sigma_{\rm T}$ and the wind velocity $V_{W}=0.01c$.

This result leads us to expect a noticeable effect from the wind in Sco X-1 where the mass supply to the disk is presumably very high. ST06 calculated the K$_{\alpha}-$ line EW  as a function of the wind optical depth $\tau_{W}$. They found  that  the change of $\tau_{W}$ from  0.2 to 1 corresponds to the EW changing from 200 eV to 1 keV. Using this ST06 result and our inferred EWs for Sco X-1 (see Figure \ref{figure4}), we conclude that the optical depth of the wind in Sco X-1 changes from 0.2 in the HB to 0.6 in the NB.  This  observational result for Sco X-1 is consistent with the TSA07 and LT04 theoretical predictions. 
 
\section{Conclusions}

A  simple model of a single Comptonizing component (CompTT) and a Iron K$_{\alpha}-$line fits all observations and all X-ray states of Sco X-1.  This modeling infers the optical depth and temperature of the CC (Fig. \ref{figure4}) and allows us to infer the photon index $\Gamma$ with small errors (Fig. \ref{figure3}). Additionally, the modeling results strongly suggest that Sco X-1 is  always in the soft state, where its photon  index  $\Gamma$  is  between 3.3 and 4.1. Cooling of the Compton cloud which suppresses efficiency of Comptonization is dictated by the strong soft emission of the disk and neutron star.  

We also demonstrate that the equivalent width (EW) of the K$_{\alpha}-$line is anticorrelated  with color ratio (NHCR)  (Figure \ref{figure4}).  The EW steadily increases as the state softens. This increase implies that the power of the wind increases,  under the assumption that the K$_{\alpha}-$line is formed in the wind. 

{\it The large optical depth of  the quasi-spherical Compton cloud in  Sco X-1 prevents us from seeing high frequency  spin pulsations from the NS}.  The apparent NS high frequency pulsations  are suppressed because of photon scattering off cloud electrons.  This leads directly to the conclusion that {\it the observed high (kHz) QPOs are only detected by Earth observers because they originate in the outer boundary of the CC}.  The position of the CC outer boundary can  be determined by the low peak of the QPO high frequency (Keplerian frequency). The NS emission as a $\sim 2.2$ keV black body radiation is never seen. 

The X-ray continuum's power decreases as the system state softens (Figure \ref{figure1}), and the apparent X-ray intensity decreases from the HB thru the NB (Figure \ref{figure2}). The source evolution shown in the color-intensity diagram also has a reduction in intensity, suggesting that the radiation pressure is always at the Eddington limit. The constant intensity in the horizontal branch results from the geometry of the Compton cloud-disk, where part of the CC low hemisphere is exposed to the observer in this branch.  In the normal branch the CC is pushed closer to NS surface  shielding the CC lower hemisphere by the disk. 

In addition, we demonstrate a definite correlation of QPOs with photon index in Sco X-1 similar to the previously reported correlation in 4U 1728-34.  The observed correlation of kHz frequencies with photon index (Figure \ref{figure3}) also leads to the conclusion that the CC contracts when the source evolves to the softer states.

\section*{Acknowledgments}

We recognize and appreciate the referee's contribution to the
discussion section.  The referee's comments helped considerably in
presenting the main results of the paper. This research has made use
of data obtained through the High Energy Astrophysics Science Archive
Research Center Online Service, provided by the NASA/Goddard Space
Flight Center. SK was partially supported by NASA Grant NAG5-12390.

\begin{appendix}

\section{The critical mass accretion rate in the disk and mass outflow rate and determining the  optical depth of the wind (outflow) }
The critical accretion rate ($\dot M^{crit}_d$) needed to launch a wind at a given disk radius R is determined by an equality of the radiation pressure and gravitational force in a given annulus between R and
R + dR and is given by  (see TSA07)

\begin{equation}
\dot M_d^{crit}(R)=\frac{8\pi}{3}R \left(\frac{H}{R}\right)\frac{m_pc}{\sigma(R)}\frac{1}{[1-(R_{\ast}/R)^{1/2}]},
\label{mdot_crit3}
\end{equation}
where $R_{\ast}$ is the central object radius (3 Schwarzchild radii for black holes and the NS radius for neutron stars); $\sigma(R)$ is an effective plasma cross section that equals the electron Thomson cross-section $\sigma_{\rm T}$, if the disk plasma is fully ionized and  $\sigma(R)>\sigma_{\rm T}$, otherwise (see e.g. Proga 2005); H is the geometrical half-width of the disk.

If $H/R$ is  constant through the disk (see Shakura \& Sunyaev  1973, hereafter SS73), then $\dot M_d^{crit}\propto R/\sigma(R)$.  In this case, the critical  mass accretion rate $\dot M_{out,d}^{crit}$  at the outer disk radius $R_{out}$  should be much higher  than  $\dot M_{in,d}^{crit}$ at the innermost  disk radius $R_{in}$, namely   
\begin{equation}
\frac{\dot M_{out,d}^{crit}}{M_{in,d}^{crit}}=\frac{R_{out}}{R_{in}}\frac{\sigma_{\rm T}}{\sigma(R_{out})}\frac{[1-(R_{\ast}/R_{in})^{1/2}]}{[1-(R_{\ast}/R_{out})^{1/2}]}.
\label{ratio}
\end{equation}
The disk works like a filter that does not allow the mass accretion rate $\dot M$ to be higher than
$M_d^{crit}(R_{in})$ at the innermost part of the disk. This implies that a much larger fraction of the total mass inflow from the companion  is diverted  to mass outflow.  This  likely  occurs in Cyg X-1, Cyg X-2 and Sco X-1 where the mass supply to the disk  is  presumably very high.
TSA07 show an expected saturation of total disk luminosity with the mass supply to the disk because 
\begin{equation}
\frac{L_{d,tot}^{crit}}{L_{d,in}^{crit}}\sim 2\times \left[\frac{\ln(R_{out}/R_{in})}{10}\right]
\left[\frac{\sigma_{\rm T}/\sigma(R_{out})}{0.2}\right].
\label{lum_ratio}
\end{equation}
Representative values for $R_{out}/R_{in}$ and  $\sigma_{\rm T}/\sigma(R_{out})$  are $\sim 10^4$ (SS73) $\sim0.2$ (Proga 2005), respectively.  The disk luminosity $L_{d,in}^{crit}$ corresponding to  the constant $\dot M_{in,d}^{crit}$  is
\begin{equation}
L_{d,in}^{crit}\sim\left(\frac{H}{R}\right)\frac{4\pi GM_{NS} m_p c}{\sigma_{\rm T}}=\left(\frac{H}{R}\right)L_{\rm Edd}.
\label{lum_crit_in}
\end{equation}
Moreover, TSA07 calculate the total mass outflow rate for sources where the mass supply from the companion is very high (near $\dot M_d^{crit}(R_{out})$).  Because 
$\dot M_d^{crit}(R)<$ $\dot M_d^{crit}(R+dR)$
(see Eq. \ref{mdot_crit3}),  the surplus  (overflow) $\Delta \dot M_d^{crit}(R)$  in the annulus 
$(R, ~R+dR)$ emerges  as outflow, i.e. the mass outflow rate is 
$\Delta \dot M_{ofl}^{crit}(R)=\Delta \dot M_d^{crit}(R)$.   Consequently, the total mass outflow rate from the disk is 
\begin{equation}
\dot M_{ofl}^{crit}=\int_{R_{in}}^{R_{out}}\Delta \dot M_d^{crit}(R)dR\sim \dot M_d^{crit}(R_{out}),
\label{outflow}
\end{equation}
which is much higher than the critical mass accretion rate at the inner disk radius
\begin{equation}
\frac{\dot M_{ofl}^{crit}}{\dot M_d^{crit}(R_{in})}\sim \frac{R_{out}}{R_{in}}\gg 1.
\label{outflow_ratio}
\end{equation}
Finally, TSA07 obtain, using Eqs. (\ref{mdot_crit3}), (\ref{outflow}) and the continuity equation, 
that Thomson optical depth of the wind (outflow) 
\begin{equation}
\tau_{W}=\frac{\dot M_{ofl}^{crit}\sigma_{\rm T}}{4\pi m_p V_{W} R_{out}}=\frac{2}{3}\frac{H}{R}
\frac{\sigma_{\rm T}}{\sigma(R)}\frac{c}{V_{W}}.
\label{tau outflow2}
\end{equation}
where $V_{W}$ is the wind (outflow)  velocity.  

\section{The Comptonization parameter  $Y$ as a reciprocal of spectral index $\alpha$}
The intensity of the injected soft photons  undergoing  $k$ scatterings in the Compton cloud is
\begin{equation}
I_k\propto p^k
\label{int_k}
\end{equation}
where $p$ is a mean probability of photon scattering in the CC. 
The probability of  photon scattering is directly related to the mean number of scatterings 
\begin{equation}
N_{sc}=\sum_{k=1}^{\infty}kp^kq=p/(1-p)
\label{scat_number}
\end{equation}
where $q=1-p$ is a probability of the photon escape from the CC.
Thus, using Eq. (\ref{scat_number}), we obtain 
\begin{equation}
p=1-1/(N_{sc}+1).
\label{p_vs_sc_number}
\end{equation}
Because the average photon energy change   per scattering  $<\Delta E> =\eta E$ (where $\eta>0$ for upscattering case),  the injected photon energy after $k$ scatterings  $E$ is
\begin{equation}
\frac{E}{E_0}=(1+\eta)^{k}.
\label{E_k}
\end{equation} 
The combination of Eqs. (\ref{int_k}),  (\ref{E_k})  yields that the emergent upscattering spectrum of the soft photon of energy $E_0$ in the bounded Compton cloud is a power law
\begin{equation}
I_{E}\propto\left(\frac{E}{E_0}\right)^{-\alpha}
\label{power_law}
\end{equation}
which which has an energy index of
\begin{equation}
\alpha=\frac{\ln(1/p)}{\ln(1+\eta)}.
\label{alpha_pl}
\end{equation}
 For $N_{sc}\gg1$ and $\eta\ll 1$ and using Eq. (\ref{p_vs_sc_number}),  we reduce Eq. \ref{alpha_pl}  to 
\begin{equation}
\alpha\approx(\eta N_{sc})^{-1}=Y^{-1}.
\label{alpha_plm}
\end{equation}
The thermal Comptonization parameter $Y\sim (4kT/m_ec^2)\tau^2$  because in this case  
$\eta=4kT/m_ec^2$ and $N_{sc}\sim \tau^2$ for $\tau\gg 1$   (see e.g. Rybicki \& Lightman 1979).  
   and, thus, the thermal Comptonization spectral index 
\begin{equation}
\alpha\sim [(4kT/m_ec^2)\tau^2]^{-1}.
\label{alpha_plmm}
\end{equation}

\end{appendix}

\clearpage


\end{document}